\documentclass{article}
\usepackage{spconf,amsmath,graphicx}
\usepackage{amsfonts}
\usepackage{enumitem}
\setlist{nosep, leftmargin=14pt}
\usepackage{tabularx, booktabs, makecell, caption}
\usepackage{mwe} 
\usepackage{xcolor}
\usepackage{parskip}
\usepackage[toc]{appendix}
\usepackage{cite}
\usepackage{capt-of,etoolbox}

\def\BibTeX{{\rm B\kern-.05em{\sc i\kern-.025em b}\kern-.08em
    T\kern-.1667em\lower.7ex\hbox{E}\kern-.125emX}}

\title{Explainable Severity ranking via pairwise n-hidden comparison:\\
A case study of Glaucoma
}

\name{%
\begin{tabular}{@{}c@{}}
Hong Nguyen $^{\star}$ \qquad 
Cuong V. Nguyen $^{\dagger}$ \qquad 
Shrikanth Narayanan $^{\star}$\\ 
Benjamin Y. Xu $^{\star \ddagger}$ \qquad 
Michael Pazzani $^{\star}$ 
\end{tabular}}
\address{
$^{\star}$Information Sciences Institute, University of Southern California\\
$^{\dagger}$College of Engineering and Computer Science, Vin University\\
$^{\ddagger}$Roski Eye Institute, Department of Ophthalmology, University of Southern California}
\begin{document}
%
\maketitle
\begin{abstract}
Primary open-angle glaucoma (POAG) is a chronic and progressive optic nerve condition that results in an acquired loss of optic nerve fibers and potential blindness.
The gradual onset of glaucoma results in patients progressively losing their vision without being consciously aware of the changes.
Accurate assessment of POAG severity is essential for timely intervention of permanent vision loss. However, ophthalmologists often disagree on severity classification, as individual thresholds for defining severity can vary. Nevertheless, they tend to reach consensus when comparing the relative severity between paired cases.
In this work, we propose a framework to compare and interpret the severity of glaucoma using fundus images using siamese-based severity ranking with pairwise n-hidden comparisons.
We additionally propose to use pair-wise sailency map to explain why a specific image is deemed more severe than others. Our findings indicate that the proposed severity ranking model surpasses traditional ones in terms of diagnostic accuracy and delivers promising saliency explanations.
\end{abstract}

\vspace{-5pt}
\section{Introduction}
\begin{figure*}[t]
\centering
  \includegraphics[width=17.5cm]{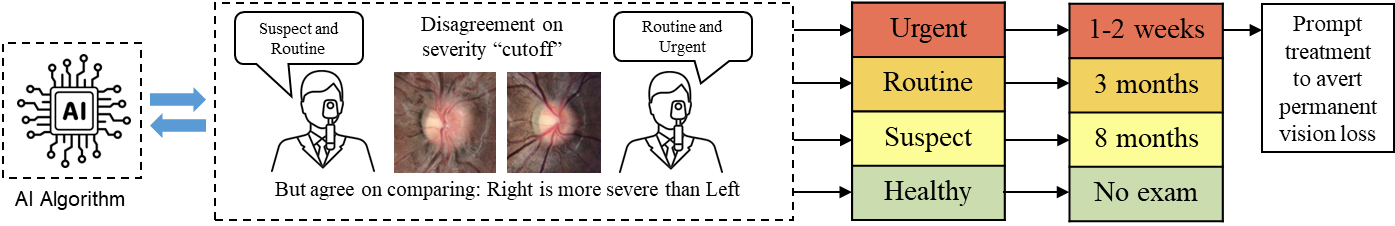}
  \caption{Paradigm Shift with AI-Enhanced Eye Care. While different ophthalmologists have different ``cut off" for severity levels, they most likely agree on comparison between pairs of image. We proposed a framework to severity ranking via pairwise comparisons.}
  \label{fig:teaser}
\end{figure*}

\vspace{-5pt}
Determining severity priority is of paramount importance in clinical settings as it serves as a critical guiding principle for healthcare professionals to allocate resources, make informed decisions, and provide timely interventions, Fig~\ref{fig:teaser}.  The accurate assessment and categorization of Glaucoma severity enables clinicians to triage patients effectively, ensuring that those in most urgent need receive immediate attention to avoid permanent vision loss.   Furthermore, comparing an earlier image to a recent image of a patient allows a clinician to determine if the disease is getting worse.
Most recent challenge on glaucoma, REFUGE \cite{ORLANDO2020101570}, focused exclusively on binary classification and segmentation tasks, without considering the aspect of disease severity.  Note that we are not advocating treating this as a four-class problem, but rather a ranking so that the cutoff between categories can be varied after learning based on resource availability.
Finally, there is often a disagreement between medical experts on ``cutoff'' of severity levels \cite{kalpathy2016plus, campbell2016plus} while they strongly agree on severity-preferred comparison between pairs.

Thanks to recent impressive advances in deep learning,  computer vision tools can now assist healthcare professionals in disease diagnosis and severity estimation from medical images. Our work investigates clinical severity ranking of medical images in the domain of ophthalmology via pair-wise comparisons. Intuitively, the model learns to decide its preference for one sample over another and subsequently ranks multiple images based on these pairwise comparisons. The decision is made based on knowledge-driven severity indicators.  For ophthalmologists, the optic cup-to-disc (CD) ratio is one of the most important indicators to identify the more severe cases of glaucoma on color fundus photography. Another indicator, the Mean Deviation index (MD-index), explicitly quantifies the vision condition of patients via comprehensive visual field testing.
We denote the CD-ratio and MD-index as ophthalmologist-centric and patient-centric severity indicators, respectively, inspired by \cite{Pazzani2022}. The correlation between the two severity indicators has been statistically investigated in the literature \cite{CDRvsMDI2017, Iutaka2017}. 
Assuming that only the signed difference between MD-index values is known, rather than the MD-index themselves, we investigate severity-prefered comparison which is ranking-based problem using pair-wise preference. It is important to note that we do not study severity classification task nor do we use any class labels.
\vspace{-5pt}
\section{Related Works}
\vspace{-5pt}
\textit{Conventional Severity Classification:} 
Numerous previous works have studied estimating disease severity from medical images using multi-class classification \cite{AIMIA22, SURI2020103960, KALPATHYCRAMER20162345} with each class signifying a distinct level of severity.
In practice, medical experts often prefer comparing pairs of samples to identify the more severe case rather than assigning each image to a predefined severity class \cite{ROP2022}.

\textit{Severity Pairwise Comparison/Ranking:} As for comparison tasks, siamese networks constitute the primary architecture choice in the literature. A siamese network is composed of two identical sub-networks with shared weights, each of which includes two consecutive modules: a feature extractor \cite{He2015DeepRL, Dosovitskiy2020AnII, Simonyan2014VeryDC} and a ranking module \cite{ltrbygd, listrank}. 
Previous studies make use of siamese networks for diverse comparison tasks, including rating urban appearance images \cite{Dubey2016DeepLT}, analyzing the ranking of burst mode shots of a scene \cite{Chang2016AutomaticTF}, and evaluating videos based on skill levels \cite{Doughty2017WhosBW}.
Most relevant to our work, Peng Tian \cite{ROP2022} compares the efficiency of comparison and classification model for the retinopathy of prematurity (ROP) dataset. Li et al.~\cite{SimSiamese2020} use multiple variants of siamese nets to evaluate continuous ROP severity. So far, the current paradigm of ranking via comparison involves a regressor that maps input features to a severity score. 
By comparing pairs of scores, it then decides which sample has a higher rank. Single-score comparison does not truly encapsulate the essence of ``comparison'' since the evaluation of severity relies on multivariate criteria and characteristics. For instance, ophthalmologists assess the condition of glaucoma based on not just CD-ratio but also on disc contour shape, vessel position, and various other factors, for arriving at the final clinical disposition.

\textit{Sailiency Explainable AI:} Several XAI algorithms \cite{lime, gradcam, shap, intergratedgrad, occlusion} have been developed to provide image-based saliency explanations. To the best of our knowledge, frameworks to apply these XAI algorithms for pairwise comparison tasks are still missing from the literature.

\textit{Contribution.} 
To address mentioned challenges and meet the intersection between comparison and explanation, our primary contributions are summarized as follows:
\begin{itemize}
  \item We propose a siamese neural network featuring n-hidden comparisons to tackle the severity comparison problem. Our results demonstrate that the siamese net with 10-hidden comparisons outperforms the state-of-the-art baseline by 11\% in terms of pairwise accuracy.
  \item We introduce novel usage of saliency map to interpret the comparisons of the proposed model. Results show different perspective on pairwise saliency explanation. 
  \item We work with opthamologists to conduct a quantitative and qualitative assessment to evaluate the comparison performance and explainability of the proposed model. 
\end{itemize}
\vspace{-3pt}
\section{Problem Statement}

Consider a training set of medical images $\{x_1,...,x_m\}$, and let $\Omega = \{(x_i, x_j) \; | \;  i,j \in \{1, ..., m \}\}$ be the collection of all possible pairs sampled from the image set, where $|\Omega| = m^2$. To reduce the burden of dimension in the training phase, we randomly select $l$ pairs from $\Omega$ as input space $\mathbb{D}_x \subseteq \Omega$ to be training set, where $|\mathbb{D}_x| = l < |\Omega|$. For each pair $(x_i, x_j) \in \mathbb{D}_x$, let $y_{i,j} \in \{0, 1\}$ denote the corresponding binary comparison label, with  $y_{i,j} = 0$ indicating that $x_i$ is less severe than $x_j$.  Note that the order of $i, j$ is important.  Let $\mathbb{D}_y = \{y_{i,j} \in \{0, 1\} \ | \  \forall\; (x_i, x_j) \in \mathbb{D}_x\}$ be the label space. The first objective is to learn the severity comparison model $f \colon {\mathbb{D}_x} \to \mathbb{D}_y$ such that $f(x_i, x_j) = y_{i,j}$. The second goal is to develop a method for interpreting the decision made by the function $f$ for any specific sample $(x_i, x_j)$ using image saliency. Specifically, the method should point out the regions of interest on both $x_i$ and $x_j$ that are important for the comparison.


\begin{figure*}[t]
\centering
  \includegraphics[width=17cm]{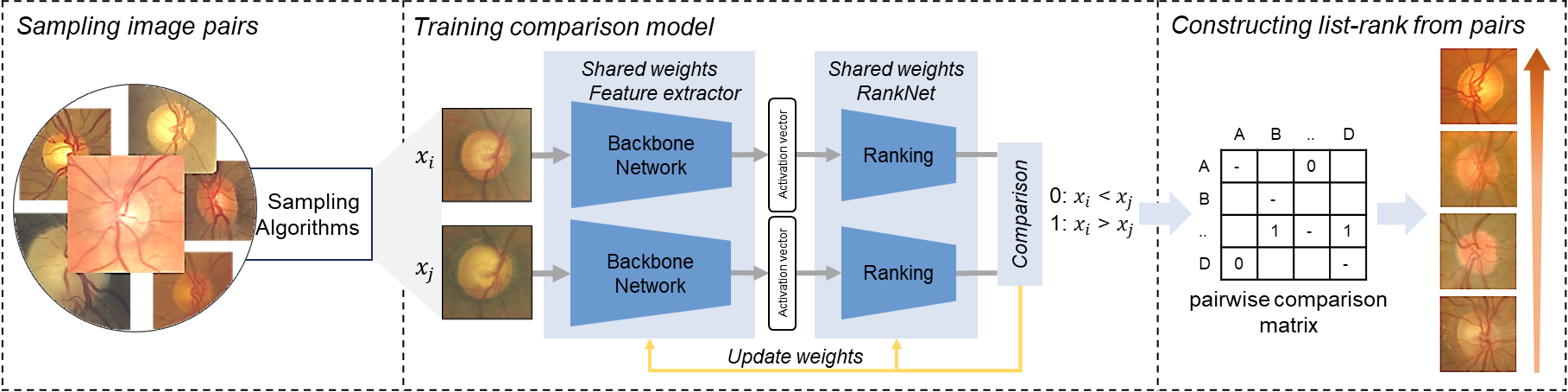}
  \caption{The three phases of our proposed framework for severity ranking via pairwise comparisons.}
  \label{framework}
\end{figure*}

\vspace{-5pt}
\section{Methodology}
The framework for severity ranking, as depicted in Figure~\ref{framework}, comprises three phases: 1) collection of comparison labels, 2) neural network training and inference, and 3) list rank reconstruction.  The labeling phase requires medical experts to provide severity annotations.
The labeled images are then partitioned into training, validation, and test sets for training and evaluating a siamese network. 
The siamese network is composed of two identical sub-networks with shared weights, each of which includes two consecutive modules: a feature extractor (backbone) and a ranking function.
We employed ResNet-50 \cite{He2015DeepRL}, VGG19 \cite{Simonyan2014VeryDC}, and Vision Transformer (ViT) \cite{Dosovitskiy2020AnII} models, pre-trained on the ImageNet dataset, as the backbone feature extractors. As for the ranking function, we customized RankNet \cite{ltrbygd}, with an input length determined by the length of activation vectors derived from the backbone.

Upon training the siamese network, we selected the best model based on validation pairwise accuracy. 
Given a pair of images, the model outputs the probability of one being more severe than the other.
Applying the model on all pairs in the input dataset yields a pairwise comparison matrix, which can be converted into a ranking list using Bradley-Terry model ~\cite{bradley1952rank}. The Bradley-Terry model is a statistical framework used for analyzing pairwise comparison data. It estimates the ranking of 
k variables, representing competitors, based on outcomes from each pairwise competition.
\vspace{-5pt}
\subsection{n-Hidden Comparison}
The conventional comparison of RankNet \cite{ltrbygd} is given by
\begin{equation}
    f(x_i, x_j) = \sigma\left(s(x_i)-s(x_j)\right),
\end{equation}
where $\sigma$ is the sigmoid activation function and $s(*) \in \mathbb{R}$ is a severity score. 
However, ranking an image may depend on many latent criteria, which might not be captured by a single score.
Therefore, we introduce a new comparison function:
\begin{equation}
    f_{\text{proposed}}(x_i, x_j) = g\left(\sigma(s_n(x_i)-s_n(s_j))\right)
\end{equation}
where $s_n(*) \in \mathbb{R}^n$ denotes a severity feature vector of size $n$, called ``n-hidden comparison", and $g(*)$ denotes a fully connected network followed by an activation function, predicting the ranking decision based on $n$ comparisons.
\vspace{-5pt}
\subsection{Explainable Framework}

Out of $n$ comparisons, we applied SHAP \cite{shap} to identify the top $k$ comparisons that contribute the most information to the final ranking decision. This XAI algorithm provides increased flexibility in selecting suitable comparison pairs. We then employed GradCAM \cite{gradcam} to generate $k$ pairs of heatmaps, before aggregating them to obtain the final explanation.
\section{Experimental Setup \& Results}
\subsection{Dataset}
The Ocular Hypertension Treatment Study (OHTS) dataset~\footnote{This research study was conducted retrospectively using human subject data made available in authorized access by
National Eye Institute. No ethical approval was required.} 
is sponsored by the National Eye Institute and collected randomly on multi-center, with a total of 1,636 subjects between 40 and 80 years of age. OHTS contains more than 74,000 fundus images of prospective treatment trials that are designed to determine intraocular pressure (IOP) and primary open-angle glaucoma (POAG). In this work, we are interested in subjects who had a POAG diagnosis. Additionally, we chose 440 patients who had multiple annual doctor visits, as their fundus images should encompass both classes, for instance, those who were initially healthy but had glaucoma the following year. Ophthalmologists assess the severity of glaucoma using the MD-index, which indicates the average increase or decrease in a patient's visual sensitivity across the entire visual field in comparison to an age-adjusted reference field of a healthy individual.
Note that the visual sensitivity test takes up to an hour for a clinician and patient while fundus photography takes less than a minute.
Given two fundus images, we determine that one is more severe than the other if its MD-index is lower.
We designed two experiment settings:

\begin{table*}[t]
  \centering
  \caption{Performance of baseline and proposed methods with respect to comparison accuracy, mean Intersection over Union (m-IoUs) and normalized discounted cumulative gain (nNCG). The baseline is a conventional model with one comparison.}
\normalsize
\begin{tabular}{lcccccccc}
\toprule
\multicolumn{1}{c}{Ranking} & \multicolumn{2}{c}{Longitudinal  Accuracy} & \multicolumn{2}{c}{Cross-sectional Accuracy} & \multicolumn{2}{c}{m-IoUs (GradCAM)} &\multicolumn{2}{c}{nDCG}\\
\cmidrule(rl){2-3} \cmidrule(rl){4-5} \cmidrule(rl){6-7} \cmidrule(rl){8-9}
  Backbone  &Baseline &10-comp.   &Baseline &10-comp. &Baseline &10-comp. &Baseline &10-comp.\\
\cmidrule(r){1-9}
VGG19 	    &\textbf{0.822}  &0.793 &\textbf{0.787} &0.753  
            &\textbf{0.023} &0.007 &0.861 &\textbf{0.927}         \\
ResNet50 	&0.759	&\textbf{0.848} &0.729 &\textbf{0.739}  
            &0.027 &\textbf{0.095} &\textbf{0.903} &0.898       \\
ViT16     	&0.800	&\textbf{0.833} &0.651 &\textbf{0.672}  
            &0.034 &\textbf{0.057} &\textbf{0.912} &0.876       \\
\bottomrule
\end{tabular}
\label{tab:compacc}
\end{table*}

\textbf{Longitudinal setting}:
We compare two images from the same subject at a time. Each image is categorized as either healthy (H) or having glaucoma (G), resulting in four possible classes for an input image pair: HH, HG, GH, and GG. We ensured that the training set was balanced across these four classes. Note that even two healthy images can be compared based on the MD-index indicator. We randomly divided the 440 selected subjects into train, validation, and test sets, with sizes of 300, 50, and 90, respectively. 
From these batches, we randomly sampled 10,000 image pairs for training and 1,000 pairs for validation and final testing.

\vspace{-5pt}
\textbf{Cross-sectional setting}:
We randomly select image pairs without considering the subjects to which they belonged. It is important to note that the dataset distribution in this case is not uniform. Additionally, when comparing image pairs from different subjects using the MD-index, there is inherent uncertainty: the closer the MD-indices of a cross-subject pair, the more uncertain the comparison becomes due to factors like noise in visual field measurements, variations in doctor preferences, or other severity indicators.
To mitigate the impact of noise and achieve a balanced dataset, we designed our sampling algorithm to choose a sample pair such that the difference between their MD-indices is greater than a predefined threshold. This threshold was set equal to half of the standard deviation of these differences. 

\vspace{-5pt}
\subsection{Results}

\textbf{n-Hidden comparison outperforms conventional comparison when activation dimension is condensed:}
As shown in Table \ref{tab:compacc}, the 10-comparison siamese network achieves a 12\% higher comparison accuracy compared to the conventional single-comparison siamese network with ResNet and ViT backbones. The main difference between VGG19 and the other backbones lies in the length of VGG19's activation vector, which is 25088, while ResNet and ViT have activation vectors with lengths of 2048 and 768, respectively. In addition to pairwise comparisons, we evaluated listwise ranking using normalized discounted cumulative gain (nDCG), which considers the positioning of relevant items within the ranked list. We computed the nDCG for each patient in the test set and took the average result. Notably, the mean nDCG of our proposed model is competitive even with an additional list-rank reconstruction phase using Bradley-Terry analysis~\cite{bradley1952rank}, outperforming the baseline with the VGG19 backbone.


\begin{figure}[t]
\centering
  \includegraphics[width=8cm]{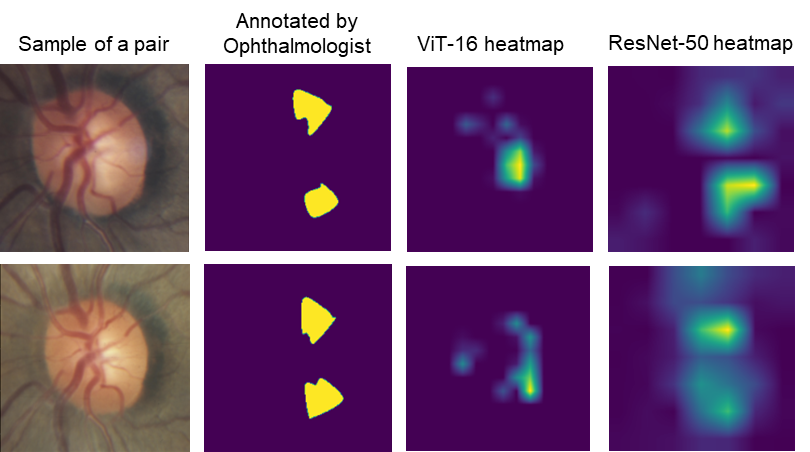}
  \caption{10-hidden comparison in best IoU case. Expert annotation show that upper image is severer than lower one because of expansion of optic nerve to upper left and lower-left region of optic disc, as present by segmentation masks.}
  \label{qualresult}
\end{figure}

\textbf{Cross-sectional comparison needs more conceptual annotations:} 
Although training on a larger set, the cross-sectional pairwise accuracy cannot meet the performance of the longitudinal one. Moreover, if we do not impose restrictions on the MD-index of pairs through our sampling method, the performance of the cross-sectional models is marginally superior to random chance. This is primarily due to the fact that cross-subject image pairs lack conceptual alignment, and the assessment of severity cannot rely solely on the MD-index. The behavior of cross-sectional comparisons mirrors that of longitude settings when changing different feature extractors. To conduct subject-specific comparisons, it is advisable to gather a more extensive set of pairwise annotations, encompassing not only the MD-index but also rim status, PSD-index, HCD-ratio, and other relevant factors.

\textbf{First step toward user-centric saliency explanations:}
For the purpose of qualitative evaluation, we ask ophthalmologists to give comparison annotations for 10 pairs of images randomly selected from the test set. Figure \ref{qualresult}(a) shows the qualitative explanation of the 10-hidden comparison siamese net with ViT-16 and ResNet-50 backbone for a sample pair. Ophthalmologists agreed that the optic nerve of the first image expands toward the upper-right and lower-right faster than the second one. Thus, the optic nerve-to-disc ratio of the first image is higher than the second. Given two ground-truth pairs of comparison, the upper-right and lower-right pair, we show two heatmaps from ViT16 and Resnet-50 that achieve the best IoUs. With the ResNet-50 backbone, the siamese model can localize well the area of interest on both images. 
On the other hand, ViT learns the concept of the rim between the optic nerve and the optic disc, which is another way of interpreting glaucoma. 

\section{Conclusions ad Future Works}
In this study, we introduce an approach aimed at enhancing the interpretability of severity ranking through a series of hidden comparisons. The results indicate that our proposed comparison method, which emphasizes a more refined dimension of feature space, outperforms conventional comparisons in terms of predictive accuracy. Along with the ranking model, we put forth an pairwise XAI usage to provide a comprehensive interpretation of ranking decisions through pairwise comparisons. We also recommend using fewer comparisons when dealing with limited annotations and acquiring more annotations for cross-sectional comparison.



\bibliographystyle{unsrt}
\bibliography{refs}

\end{document}